\documentclass[12pt]{iopart}

\usepackage{graphicx}
\usepackage{color}

\renewcommand{\vec}[1]{\mbox{\boldmath$\mathrm{#1}$}}
\def\ind#1{{_{\mathrm{#1}}}}

\begin{document}

\title[Beating the superparamagnetism]{Beating  the superparamagnetic size limit of nanoparticles on a ferroelectric substrate}

\author{A. Sukhov$^1$, L. Chotorlishvili$^1$, P.P. Horley$^2$, C.-L. Jia$^3$, S.K. Mishra$^4$, J. Berakdar$^1$}
\address{$^1$Institut f\"ur Physik, Martin-Luther Universit\"at Halle-Wittenberg, 06120 Halle (Saale), Germany}
\address{$^2$Centro de Investigaci\'{o}n en Materiales Avanzados (CIMAV S.C.), Chihuahua/Monterrey, 31109 Chihuahua, Mexico}
\address{$^3$Key Laboratory for Magnetism and Magnetic Materials of the MOE, Lanzhou University, Lanzhou 730000, China}
\address{$^4$Department of Applied Physics, Indian Institute of Technology, Banaras Hindu University, Varanasi - 221005, India}
\ead{alexander.sukhov@physik.uni-halle.de}
%
%

\begin{abstract}
When decreasing the size of nanoscale magnetic particles their magnetization becomes vulnerable to thermal
fluctuations as approaching  the superparamgnetic limit, hindering thus applications relying on a stable magnetization. Here, we show theoretically that  a  magnetoelectric coupling to a ferroelectric
substrate renders possible the realization of  substantially smaller nano clusters  with thermally stable magnetization. For an estimate of cluster size we perform calculations with realistic material parameters for iron nano particles on ferroelectric BaTiO$_3$ substrate. We find, steering the polarization of BaTiO$_3$ with  electric fields affects the magnetism of the deposited magnetic clusters. These findings point to a qualitatively new class of  superparamagnetic composites.
\end{abstract}


\maketitle
\section{Introduction}
Fueled by novel fabrication, miniaturization, and characterization techniques nanomagnetism has been a crucial element for the ongoing advances in nanotechnology. E.g., the current  high-density information magnetic storage is achieved by nano structures  having out-of-plane (or perpendicular) magnetization \cite{WeMo99}. Decreasing further the size of the nanostructures would allow for yet higher storage capacity, but at some critical size (depending on the material, typically below 3-50 nm)  the magnetization starts  flipping randomly its  direction  due to thermal activation which marks superparamagnetic state of the system \cite{Neel49}. This phenomenon is well-known with implications for a variety of applications including  magnetic fluids, magnetic refrigeration, magnetic resonance imaging, and drug delivery schemes \cite{LaDu11,GoGa13,WeVa10}. Thus, finding ways to tune superparamagnetism and/or to stabilize the magnetization while downsizing the nano clusters is an issue of key importance. A highly desirable way would be to achieve this goal via external electric means which would imply less energy consumption than magnetic fields and more flexibility than the synthesis  of new material compositions with engineered magnetic anisotropy energy density $K$ which is a decisive factor for superparamagnetism. Indeed the N{\'e}el relaxation time $\tau_N$, i.e. the mean time between successive  flips depends exponentially on the magnetic anisotropy energy $KV$ ($V$ is the particle volume) with respect to the thermal energy \cite{Neel49} (i.e. $\tau_N = \tau_0 \exp \left(\frac{K V}{k_{\mathrm{B}} T}\right)$, where $\tau_0$ is a material specific  attempt time). Small variations in the energy barrier  $KV$ affect substantially $\tau_N$. In particular, increasing $K$ allows for smaller $V$ while maintaining  $\tau_N$. From this perspective, coupled ferroelectric/magnetic nanostructures \cite{Fieb05,EeMa06,RaSp07,NaBi08} such as those shown in Fig. \ref{fig_1} are highly interesting: Indeed, studying the
ferromagnetic resonance behavior of such a composite material we have shown recently  {theoretically} \cite{SuHo13} that the magnetoelectric (ME) coupling acts as an additional {unidirectional} anisotropy that can be influenced by an electric field due to the electric response of the ferroelectric part of the composite structure.
These predictions were subsequently confirmed  in a recent experiment for Co/BaTiO$_3$-interface \cite{JeBa13}.
Superparamagnetic behavior of this structure is expected thus to be highly sensitive to the   magnetoelectric coupling, even if this coupling is weak.
 It is worthwhile noting that sofar, a major obstacle for applications of multiferroics is the smallness of the magnetoelectric coupling. Hence, 
 an efficient electric field control of magnetism is possible only in close proximity of the interface. In the present work
 we exploit two advantages: 1) we consider nano systems where the aforementioned  proximity is a major part of the whole object, and 2) we concentrate on a phenomena that
is influenced by the magnetoelectric coupling  in an exponential manner (because ME-coupling acts
an additional  anisotropy   \cite{SuHo13,JeBa13}),
 circumventing  thus the obstacle of a small ME coupling.\\
  The purpose of the current study is to formulate the  posed problem rigorously and to conduct numerical simulations to quantify the above statements and expectations.
  To this end, and to be specific we consider Fe nanoclusters deposited on a ferroelectric  substrate (BaTiO$_3$), cf. Fig. \ref{fig_1}. \\
  Our analysis and numerical simulations  confirm the above expectation that superparamagnetism in this composite structure can be controlled electrically and the size limit can be pushed down substantially. Our theory is based on the Fokker-Planck equation \cite{Fokk14,Risk89} constructed for the Landau-Lifshitz-Gilbert equation \cite{LaLi35,Gilb55} and yields the Arrhenius-like exponential behavior for the N{\'e}el relaxation time $\tau_N$ of the magnetization. Further numerical simulations on the thermodynamic properties for the coupled polarization and magnetization dynamics revealed an intimate relationship between the  magnetoelectric coupling strength and the superparamagnetic behavior.

\section{Theoretical model}
\begin{center}
   \begin{figure}[htb]
    \centering
    \includegraphics[width=.8\textwidth]{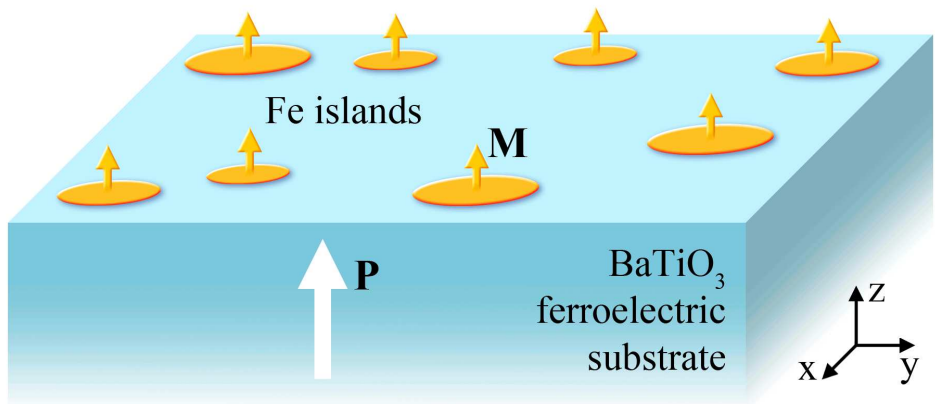}
        \caption{\label{fig_1} (Color online) Schematics of the studied composite structure: single-domain magnetic  nanoparticles are deposited or immersed in a ferroelectric substrate. Coupling between the {interface polarization $P_1$} and the magnetization $M$ allows for shifting and electric tuning of the superparamagnetism of the nanoparticles.}
    \end{figure}
\end{center}
The treatment of the coupling of a system to a thermal bath is an established case study for  experiments and  theory. In experiments the
thermal properties might be captured e.g. by the telegraph noise technique\cite{CaSn91,WeOr97}, where the response of the system is measured as a
function of time at different temperatures. For a two-state system the number of switchings between the two stable states can be counted and the average
of these switches yields a mean switching time at a given temperature. Theoretically, the physical process might be formulated in the framework of
the Langevin approach\cite{Lang08}, for which the properties of thermal (white) noise might also be defined. From the construction of the corresponding
Fokker-Planck equation \cite{Fokk14,Risk89} the so called Mean First Passage Time (MFPT) - the average time needed for the system to overcome the potential
barrier due to thermal fluctuations - can be defined\cite{Risk89,HaTa90,EnMo88} and calculated. The MFPT is inseparably associated with the experimental
switching time of the system.
The general theory was applied by N{\'e}el \cite{Neel49} and subsequently by Brown \cite{Brow63} to small ferromagnetic single-domain grains/nanoparticles.
During the last decades these theoretical findings were endorsed by experiments performed at low temperatures in the presence of static magnetic
fields \cite{WeOr97}, ac-magnetic fields\cite{ThWe02,ThWe03} or spin-polarized currents \cite{KrBe07,KrHe09}.  {In the present study} we focus on the composite multiferroic structure, depicted in Fig. \ref{fig_1}, which corresponds to a 2-0 laminate multiferroic system in the terminology of Ref. \cite{VaHo10}.
  To be specific we perform concrete simulations of the ferroelectric (FE) BaTiO$_3$ substrate
  and ferromagnetic (FM) Fe nanoparticles. The density of these nanoparticles
  is chosen as low enough such that inter-nanoparticle couplings can be ignored. For this system it has been shown theoretically
   {\cite{CaJu09,LeSa10,ScBo12}} and demonstrated experimentally \cite{GaBi10,MeKl11} that the magnetoelectric coupling originating
from the spin-polarized screening charges at the FE/FM interface \cite{Zhan99} is large and stable even at room temperature \cite{LeSa10}.  {In view of the already realized experiment {\cite{GoHa13}} it is well conceivable that the suggested system in Fig. \ref{fig_1} is readily realizable by e.g. increasing the annealing temperature for 26 monolayers of Fe on the BaTiO$_3$(100)-substrate which results in the formation of Fe nanoclusters on BaTiO$_3$ with sizes from 30 to 150 nanometers. Construction of an analytical model based on the Fokker-Planck equation for FM nanoparticles in multi-domain state is quite involved. Here we  focus on small FM nanoparticles, where a single-domain state is
prevalent.} \\
\emph{Treatment of thermal effects in ferromagnets and ferroelectrics}.-\\
The treatment of thermal fluctuations in ferromagnets is well established\cite{Brow63,Kubo6370,Gara97,GaLa98,CoKa12}, and here we extend the approach as to
include the influence of the magnetoelectric coupling. The starting point is the Landau-Lifshitz-Gilbert (LLG) equation\cite{LaLi35,Gilb55}   describing the classical dynamics of the magnetization $\vec{M}(t)$ of a FM nanoparticle
\begin{equation}
    \displaystyle \frac{d\vec{M}}{dt} = - \frac{\gamma}{1+\alpha^2_{\mathrm{FM}}} \left[\vec{M}\times \vec{H}^{\mathrm{FM}}(t)\right] - \frac{\alpha\ind{FM} \gamma}{1+\alpha^2_{\mathrm{FM}}}\frac{1}{M\ind{S}} \left[\vec{M}\times\left[\vec{M}\times\vec{H}^{\mathrm{FM}}(t)\right]\right],
\label{eq_1}
\end{equation}
where  $\alpha\ind{FM}$ is a (Gilbert) damping parameter \cite{Gilb55}.
The total effective field $\vec{H}^{\mathrm{FM}}(t)$ is  a sum of deterministic and stochastic contributions\cite{footnote_1}
\begin{equation}
    \displaystyle{\vec{H}^{\mathrm{FM}}(t) = - \frac{\delta F\ind{\Sigma}}{\delta \vec{M}}}+\vec{\zeta}(t),
\label{eq_2}
\end{equation}
where $$F\ind{\Sigma}=F_{\mathrm{FE}}+F_{\mathrm{FM}}+E_{\mathrm{C}}/V_{\mathrm{C}}$$ is the total free energy density of the system
containing the FM ($F_{\mathrm{FM}}$) and the FE ($F_{\mathrm{FE}}$) contributions as well as the term  $E_{\mathrm{C}}/V_{\mathrm{C}}$ due to the ME coupling.
Note because of the latter term $\vec{H}^{\mathrm{FM}}(t)$ depends also on the polarization.
 In eq. (\ref{eq_2}) $\vec{\zeta}(t)$ is an
effective stochastic field originating from the coupling of the magnetization to the thermal bath  and $V_{\mathrm{C}}$ is the volume relevant for FE/FM coupling. The thermal noise is usually assumed as a white Gaussian noise with the following characteristics\cite{LyBe93,LyCh93,Nowa01}
\begin{equation}
\displaystyle \langle \zeta_i(t)\rangle =0, \,\,\,\, \langle \zeta_i(t)\zeta_j(t')\rangle = \frac{2\alpha\ind{FM}k_{\mathrm{B}}T}{\gamma M\ind{S} V_{\mathrm{FM}}} \delta_{ij}\delta(t-t'),
\label{eq_3}
\end{equation}
where indexes $i$ and $j$ stand for different cartesian components, $\gamma$ denotes the gyromagnetic ratio, $M\ind{S}$ stands for the saturation
magnetization and $V\ind{FM}$ is the volume of the FM system/nanoparticle.\\
Treatment of thermal effects in  FE materials  may be conducted along several lines \cite{StLe98,RaAh07,LiCr05}. For FE it is widely accepted that temperature affects the coefficients of  the Ginzburg-Landau-Devonshire (GLD) \cite{Ginz49,Devo49} potential, $F_{\mathrm{FE}}$, yielding thus an effective lowering of the potential barrier when elevating the temperatures. This also accounts for modifications of the GLD-potential due to structural changes of FEs \cite{RaAh07}.\\
Attempts to introduce the thermal noise into the dynamics of FEs in a similar way as has been done for FMs exist \cite{AuKl03,SiWi04}. In this case the deterministic Landau-Khalatnikov (LKh) equation for the relaxation of the coarse-grained discrete polarization $\vec{P}_k(t)$ \cite{LaKh54}
\begin{equation}
\displaystyle {\frac{d\vec{P}_k}{dt} = -\frac{1}{\gamma_{\nu}}\frac{\delta F_{\mathrm{\Sigma}}}{\delta \vec{P}_k} + \frac{1}{\gamma_{\nu}} \vec{E}_{k \mathrm{th}}(t)}
\label{eq_4}
\end{equation}
is supplemented by a fluctuating electric field $\vec{E}\ind{th}(t)$ in full analogy with the thermal field for the FM (eq. (\ref{eq_3})) \cite{SiWi04}
\begin{equation}
\displaystyle {\langle E_{k i} \ind{th}(t)\rangle =0, \,\,\,\, \langle E_{k i} \ind{th}(t)E_{k j} \ind{th}(t')\rangle = \frac{2\gamma_{\nu}k_{\mathrm{B}}T}{V _{\mathrm{FE}}} \delta_{ij}\delta(t-t')},
\label{eq_5}
\end{equation}
whereby $\gamma\ind{\nu}$ is the FE relaxation constant and $V_{\mathrm{FE}}$ is the volume of the FE. $i$ and $j$ index the cartesian components of the $k$-th polarization $\vec{P}_k(t)$.\\
The Fokker-Planck equations can be constructed for the LLG eq. (\ref{eq_1}) with definitions (\ref{eq_2}) and (\ref{eq_3}) and separately for the LKh eq. (\ref{eq_4}) with the characteristics (\ref{eq_5}). Upon introducing a probability density current\cite{Risk89} (p. 96) and in the approximation of high potential barrier as well as absorbing boundary conditions the FM or FE MFPT can be calculated.

\emph{Zero field thermal stability of ferromagnets and ferroelectrics}.-
As depicted in Fig. \ref{fig_1} the  FE substrate is macroscopically large and possesses a stable polarization $P\equiv P\ind{z}$, which is aligned
along the $z$ axis (tetragonal phase for BaTiO$_3$\cite{Merz49}, for a temperature range from 273 [K] to 392 [K]).
The Fe nanoparticles are of sizes such that they are at the verge of superparamagnetic limit (or below) and their density is such that no interactions
between them need to be considered. Hence, it suffices to consider the dynamics of one of these nano clusters.
To be able to compare the numerical results with analytical findings the FM nanoparticles should be in a single-domain state and the magnetization switching
should occur in a collinear manner. For a FM nanoparticle to be in a single-domain state contributions from  {the exchange, the magnetocrystalline anisotropy and the FM dipole-dipole interactions should be compared \cite{Coey10} (p. 267)}. Coherent magnetization switching takes place for the sizes when the exchange interaction exceeds the dipolar interactions (exchange length). To fulfill the above restrictions the diameter of iron nanoparticle should be
around $d\ind{FM}=10$ [nm] \cite{Coey10}. For these sizes probable contributions from the surface anisotropy are rather small \cite{Fior05}.\\
 To address the issue of the magnetization stability  of the Fe nanoparticles
  against  thermal fluctuations we  calculate the MFPTs. It is known \cite{CoKa12}, that for zero magnetic field the FM MFPT is given by equation
\begin{equation}
\displaystyle \tau^{\mathrm{MFPT}}_{\mathrm{FM}}=\frac{\tau\ind{N}\sqrt{\pi}}{\left(\frac{K_1V\ind{FM}}{k\ind{B}T}\right)^{3/2}}\cdot \mathrm{e}^{\frac{K_1V\ind{FM}}{k\ind{B}T}},
\label{eq_6}
\end{equation}
where $K_1(\mathrm{Fe})=4.8\cdot 10⁴$ [J/m$^3$]\cite{Coey10} is the magnetocrystalline anisotropy strength, the FM volume is $V_{\mathrm{FM}}=\pi/6 d^3_{\mathrm{FM}}$ and the free diffusion time is defined  $\tau\ind{N}=\frac{V\ind{FM}M\ind{S}(1+\alpha^2_{\mathrm{FM}})}{2\gamma \alpha\ind{FM}k\ind{B}T}$\cite{CoKa12}.  {The value of the damping parameter is $\alpha_{\mathrm{FM}}=0.02$ \cite{CeHe91}}. The saturation of the Fe magnetization is $M_{\mathrm{S}}=1.71\cdot 10^6$ [A/m]\cite{Coey10}.\\
For iron nanoparticle with the given constants and volume at $T=293$~[K] we have
\begin{equation}
\displaystyle \tau^{\mathrm{MFPT}}_{\mathrm{FM}}(d_{\mathrm{FM}}=10~[\mathrm{nm}]) \approx 9\cdot 10^{-7}~[\mathrm{s}].
\label{eq_7}
\end{equation}
The result of the MFPT derived from the Fokker-Planck equation constructed for an overdamped one-dimensional equation of motion\cite{Risk89,HaTa90} can easily be applied to eq. (\ref{eq_4}), which has the same form for the $z$-component of the polarization. At these conditions ($T=293$~[K], $V\ind{FE}=d_{\mathrm{FE}}\times l \times l$, where $d_{\mathrm{FE}}$ is the thickness of the FE and $l \times l$ is its cross section) and in the high barrier limit for the FE potential of BaTiO$_3$ ($|\alpha_{\mathrm{FE}}|=2.77\cdot 10^7$ [Vm/C]\cite{HlMa06}, $\beta\ind{\mathrm{FE}}=1.7\cdot 10^8$ [Vm$^5$/C$^3$]\cite{HlMa06})
\begin{equation}
\displaystyle F_{\mathrm{FE}}= - \frac{\alpha_{\mathrm{FE}}}{2}P^2 + \frac{\beta_{\mathrm{FE}}}{4}P^4,
\label{eq_8}
\end{equation}
the zero-electric-field FE MFPT can be estimated as ($d_{\mathrm{FE}}=10 \mathrm{[nm]}$, $l=5 \mathrm{[nm]}$ and $\gamma_{\nu}=1.5\cdot 10^{-5}$ [Vms/C]\cite{Hlin07})
\begin{equation}
\displaystyle \tau^{\mathrm{MFPT}}_{\mathrm{FE}}=\frac{\pi}{\sqrt{2}}\frac{\gamma_{\nu}}{\alpha\ind{FE}}\mathrm{e}^{\frac{\alpha^2_{\mathrm{FE}}V\ind{FE}}{4\beta k\ind{B}T}} \approx 4\cdot 10^{18}~[\mathrm{s}].
\label{eq_9}
\end{equation}
The comparison of eqs. (\ref{eq_7}) and (\ref{eq_9}) clearly demonstrates the stability of the FE layer against  thermal noise,
meaning that on the time scale of the experiment the FE substrate does not   switch upon  thermal noise.\\
In real experiments, however,  FE  might need to be attached  to metallic electrodes  which results in depolarizing
fields\cite{MeSi73,PeKo07,RaAh07,BrLe09}. The FE barrier constant should then be modified according to\cite{RaAh07}
\begin{equation}
\displaystyle \tilde{\alpha}\ind{FE} = - |\alpha\ind{FE}| + \frac{2 d\ind{int}}{d\ind{FE}\varepsilon_0 \varepsilon\ind{int}},
\label{eq_10}
\end{equation}
where $d_{\mathrm{int}}\approx 1$ [nm] is the thickness of the so called "dead" layer (screening length)\cite{BrLe09} and $\varepsilon_{\mathrm{int}}=100$ is its dielectric constant.\\
Such small modifications of $\alpha\ind{FE}$ can significantly change the FE MFPT. Thus, for $d_{\mathrm{FE}}=100$ [nm] and $l=5$ [nm] we find for
the FE MFPT from eq. (\ref{eq_9}) with eq. (\ref{eq_10}) $\tau^{\mathrm{MFPT}}_{\mathrm{FE}}\approx 1$ [s].\\
Comparing  eqs. (\ref{eq_7}) and  FE MFPT estimate including depolarizing fields we may safely
say that the ferroelectric substrate is not perturbed by heat and we do not
need to consider thermal fluctuations in  eq. (\ref{eq_4}).
\section{Numerical simulations}
\begin{center}
   \begin{figure}[!t]
    \vspace{-5.ex}
    \includegraphics[width=.7\textwidth,angle=-90]{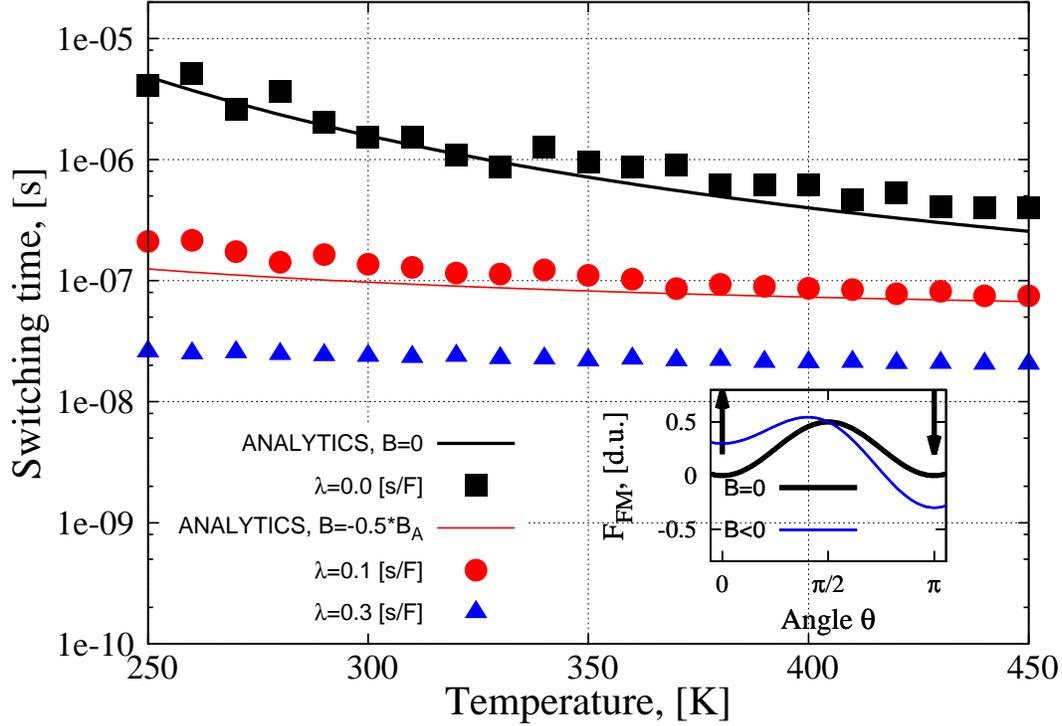}
        \caption{\label{fig_2} (Color online) Influence of the strength of the magnetoelectric coupling $\lambda$ on the  {averaged} switching time $\tau^{\mathrm{Sw}}_{\mathrm{FM}}$ of the magnetization. Diameter of the FM nanoparticle is $d_{\mathrm{FM}}=10$ [nm], $d_{\mathrm{FE}}=10$ [nm]. Points represent numerical simulations, whereas the colored curves follow from eq. (\ref{eq_11}) with the effective field taken as $b\sim \lambda P_{\mathrm{z}}/B_{\mathrm{A}}$, where $B_{\mathrm{A}}=2K_1/M_{\mathrm{S}}$. No external electric or magnetic fields are applied. Due to computational limitations the maximum length of the time scale  {(related to experimental measurement time)} is set to $20$~[$\mu$s]. Other parameters are listed in the text. Inset schematically shows the initial magnetization state (arrow up) and the change of the free energy $F_{\mathrm{FM}}$ for non-zero ME coupling $\lambda$, when polarization $\vec{P}_1$ and magnetization $\vec{M}$ are parallel to each other.}
\end{figure}
\end{center}
\begin{center}
   \begin{figure}[htb]
    \centering
    \includegraphics[width=.65\textwidth,angle=-90]{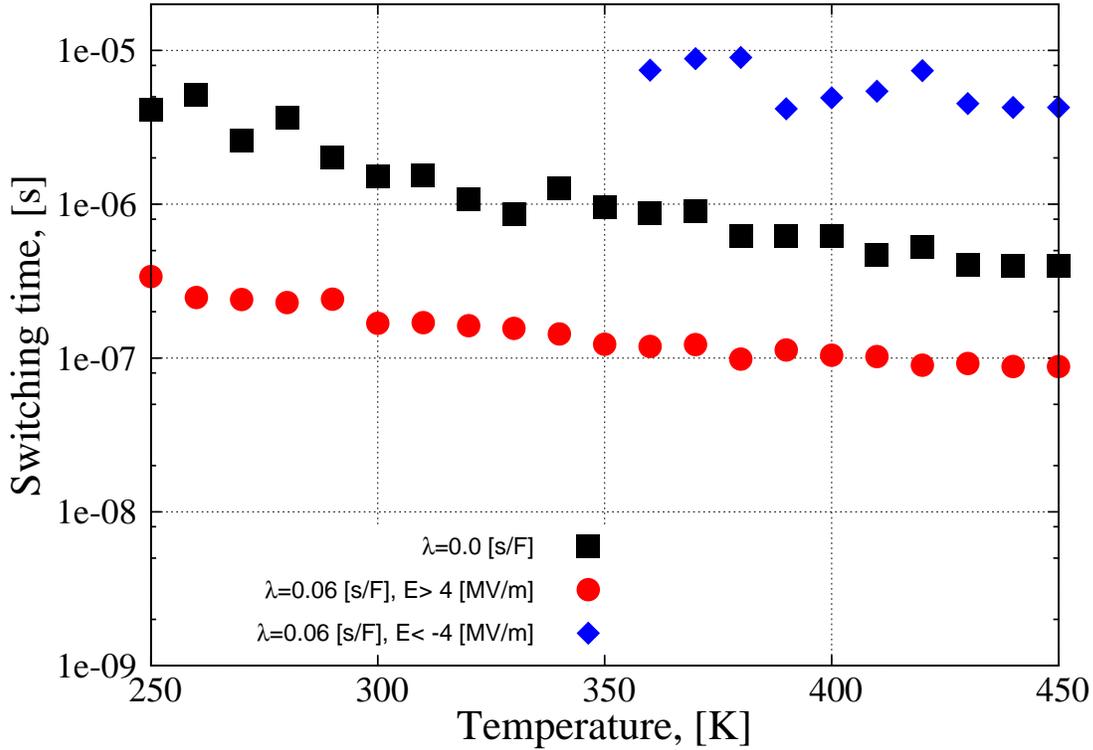}
        \caption{\label{fig_3} (Color online) Influence of the direction of the applied electric field on the  {averaged} switching time $\tau^{\mathrm{Sw}}_{\mathrm{FM}}$ of the magnetization. For each point obtained numerically the initial state of the magnetization is fixed as $\vec{M}(t=0)=M_{\mathrm{z}}\vec{e}_{\mathrm{z}}$. ME coupling was chosen to correspond the realistic value for BaTiO$_3$/Fe-junction, $\lambda=0.06$ [s/F]\cite{HoSu12}, $d_{\mathrm{FM}}=10$ [nm], $d_{\mathrm{FE}}=10$ [nm].}
\end{figure}
\end{center}
\begin{center}
   \begin{figure}[htb]
    \centering
    \includegraphics[width=.65\textwidth,angle=-90]{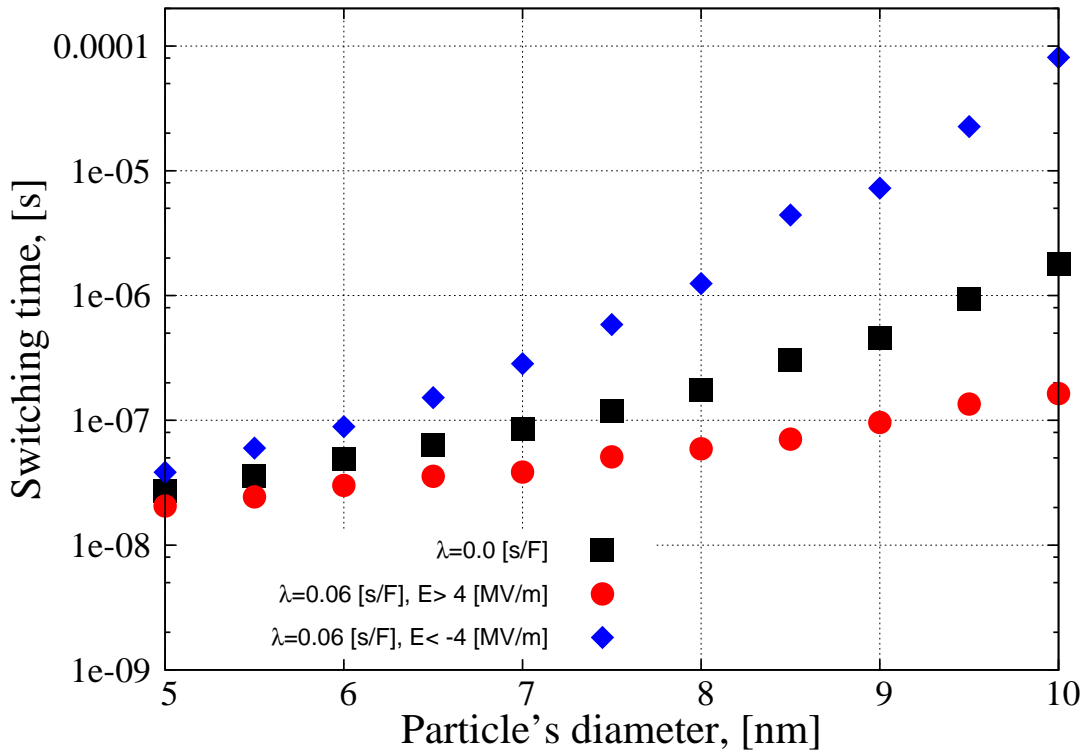}
        \caption{\label{fig_4} (Color online) Demonstration of the size effect on the  {averaged} switching time  $\tau^{\mathrm{Sw}}_{\mathrm{FM}}$ of the magnetization at room temperature $T=300$ [K]. For each point obtained numerically the initial state of the magnetization is chosen according to $\vec{M}(t=0)=M_{\mathrm{z}}\vec{e}_{\mathrm{z}}$. Further parameters are as those listed in Fig. \ref{fig_3}.}
    \end{figure}
\end{center}
\begin{center}
   \begin{figure}[htb]
    \centering
    \includegraphics[width=.65\textwidth,angle=-90]{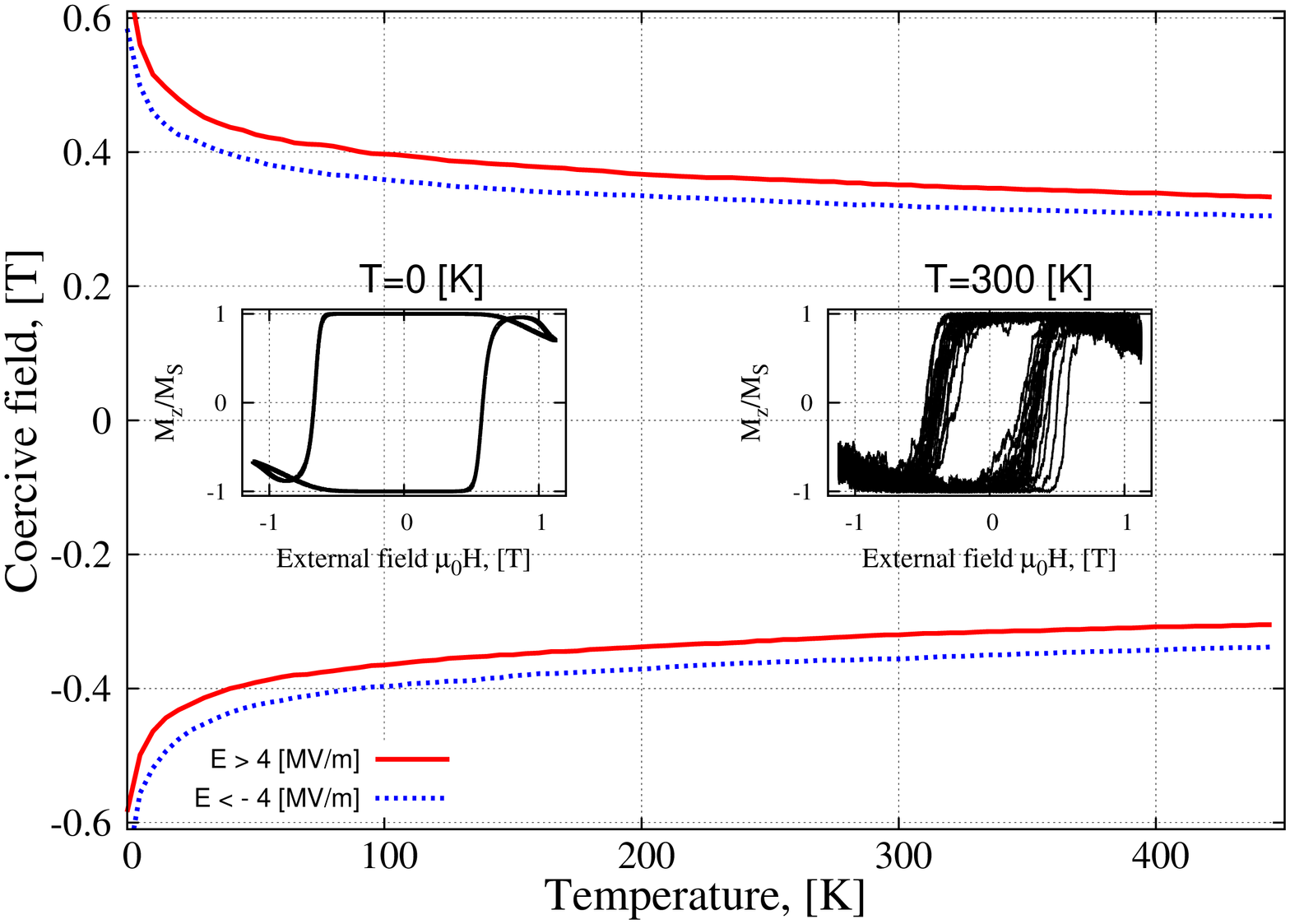}
        \caption{\label{fig_5}  {(Color online) Dependence of coercive fields on temperature and the direction of the applied electric field, calculated for a single iron-nanoparticle on the BaTiO$_3$ single-crystal substrate. The averaging was performed for ten last periods of the external magnetic field. Insets demonstrate hysteresis curves for zero Kelvin and for $T=300$~[K] without averaging (as-calculated data) as a response to external magnetic field $B_{\mathrm{z}}=\mu_0 H \cos \omega t$, where $\omega/(2\pi)=0.01$~[GHz]. The ME coupling is $\lambda=0.06$~[s/F] and further parameters are as those listed in Fig. \ref{fig_3}.}}
    \end{figure}
\end{center}

From the theory of magnetization dynamics at finite temperatures \cite{Brow63} the FM MFPT at finite applied static magnetic fields is known. The switching time, which is defined as $\tau^{\mathrm{Sw}}_{\mathrm{FM}}=2\left(\frac{1}{2\tau^{\mathrm{MFPT}}_{\uparrow}}+\frac{1}{2\tau^{\mathrm{MFPT}}_{\downarrow}}\right)^{-1}$ reads \cite{CoKa12} then in the presence of a reduced static magnetic field $b=-\frac{1}{(2K_1/M_{\mathrm{S}})}\frac{\delta F_{\Sigma}}{\delta M_{\mathrm{z}}}$
\begin{equation}
\displaystyle \frac{\tau^{\mathrm{Sw}}_{\mathrm{FM}}}{2}=\frac{\tau\ind{N}\sqrt{\pi}}{\left(\frac{K_1V\ind{FM}}{k\ind{B}T}\right)^{3/2}(1-b^2)} \cdot \frac{1}{(1+b)\mathrm{e}^{-\frac{K_1V\ind{FM}}{k\ind{B}T}(1+b)^2}+(1-b)
\mathrm{e}^{-\frac{K_1V\ind{FM}}{k\ind{B}T}(1-b)^2}}.
\label{eq_11}
\end{equation}
Here times $\tau^{\mathrm{MFPT}}_{\uparrow}$ and $\tau^{\mathrm{MFPT}}_{\downarrow}$ denote the MFPTs for the classical magnetization $\vec{M}(t)$ depending on the orientations of the applied magnetic field and the initial magnetization $\vec{M}(t=0)$.
 As discussed in previous studies  \cite{SuHo13,CaJu09,SuJi10,HoSu12} the magnetoelectric coupling  at the  FE/FM interface is due to a spin-polarized charge rearrangement (screening) and can be modeled by
   {$E_{\mathrm{C}}/(V_{\mathrm{C}})=\lambda \vec{P}_1\cdot \vec{M}$}, where $\lambda $ is a pseudoscalar characterizing the strength of this coupling, and $\vec{P}_1$ is the polarization at the FM interface \cite{footnote_3}. The FM system is considered as a coherently rotating macroscopic magnetization, the dynamics of which is governed by eqs. (\ref{eq_1}) and (\ref{eq_2}) with the inclusion of thermal noise  according to eq. (\ref{eq_3}). We have
\begin{equation}
\displaystyle F\ind{FM}=-\frac{K_1}{M^2_{\mathrm{S}}}M^2_{\mathrm{z}}-\vec{B}(t)\cdot \vec{M},
\label{eq_12}
\end{equation}
containing magnetocrystalline anisotropy aligned along the $z$-axis and the Zeeman interaction (when $B(t)\neq 0$). The resulting stochastic LLG equation that
also includes implicitly  the polarization dynamics and the ME coupling through  the effective field (\ref{eq_2}), is numerically solved using the Heun method, which converges in quadratic mean to the solution interpreted in the sense of Stratonovich \cite{Kamp07}. The dynamics of the FE spatially discretized polarization vector $\vec{P}_k(t)$ is governed by eq. (\ref{eq_4}) with zero thermal noise, however. The FE free energy in this case reads
\begin{equation}
\displaystyle F\ind{FE}  = \sum_k \left(- \frac{\alpha_{\mathrm{FE}}}{2}P^2_k+\frac{\beta_{\mathrm{FE}}}{4}P^4_k+\frac{\kappa_{\mathrm{FE}}}{2}\left(P_k-P_{k-1}\right)^2 \right) - E_{\mathrm{z}}P_{\mathrm{tot}},
\label{eq_13}
\end{equation}
with $E_{\mathrm{z}}$ and $P_{\mathrm{tot}}\equiv P$ being the $z$-component of the static electric field and the total polarization, respectively. Values of the coefficients entering the FE potential (\ref{eq_13}) are known\cite{SuJi10}. The numerical solution of the coupled equations (\ref{eq_1}) and (\ref{eq_4}) proceeds as described in Ref. \cite{SuJi10}. Details of numerical definitions and realizations of switching times of the magnetization can be found in Ref. \cite{SuBe08}. We note, however, that the tests performed for FM nanoparticles only ($\lambda=0$) are in full quantitative and qualitative agreement with the analytical results inferred  from eqs. (\ref{eq_6}) and (\ref{eq_11}).\\
We first inspect the influence of the strength of ME coupling on the behavior of FM switching times (Fig. \ref{fig_2}). The switching times are presented as a  function of temperature, which is one of the natural external parameters assisting the switching. Because of the parallel alignment of the magnetization and the polarization, increasing $\lambda$ results in negative effective magnetic field $-\lambda P_{\mathrm{z}}$ which elevates the initial state of $\vec{M}$ and hence lowers the overall switching time (inset of Fig. \ref{fig_2}). In addition, switching times obtained numerically and from eq. (\ref{eq_11}) show a good agreement in the chosen range of applied fields and temperatures, meaning the possibility of fitting of experimental data with analytical expressions, e.g. based on eq. (\ref{eq_11}).\\
Now we consider the case of a finite ME coupling strength  $\lambda=0.06$ [s/F]\cite{HoSu12}; this value  is based on the \textit{ab-initio} calculations. We find that the FM switching times are sensitive to the applied electric field $E_{\mathrm{z}}$ (Fig. \ref{fig_3}), however, not to the strength of the field, but rather to its direction, i.e., once the strong electric field switched the polarization (red  and blue points in Fig. \ref{fig_3}), it also led to visible changes of the FM switching times. \\
An essential finding depicted in Fig. \ref{fig_4} is that for a fixed
 particle size the switching time can be modified by more than an order of magnitude by the ME coupling assisted by an electric field.
In other words, the size of particles with the same switching time can be decreased substantially by ME coupling. In fact, particles that are already superparamagnetic can be stabilized by  ME coupling. This fact can in turn serve as an indicator for ME coupling, in particular, the dependence of the applied external electric field. As stated in the introduction this strong dependence is not really surprising since the switching time depends exponentially on the energy barrier that is modified in a linear way by the ME coupling (cf. Ref. \cite{Risk89} for a discussion of the physics behind the exponential behavior).\\
 {In addition to switching times the calculations of hysteresis were performed for the temperature range accounting for different structural phases of BaTiO$_3$ (Fig. \ref{fig_5}). Since temperature assists the magnetization switching process, it results in a reduction of the coercive fields. Relatively low frequency $\omega/(2\pi)=0.01$~[GHz] corresponding to the period of $T=100$~[ns] was chosen to provide sufficient relaxation, whereas the relaxation time of the nanoparticle is mainly governed by the damping parameter and scales according to $T_{\mathrm{Rel}}\approx T_{\mathrm{A}}/\alpha_{\mathrm{FM}}=\frac{2\pi M_{\mathrm{S}}}{\gamma 2K_1 \alpha_{\mathrm{FM}}}\approx 33$~[ns]. Additionally, we observe a small shift of hysteresis curves due to the presence of a unidirectional anisotropy induced by the ME coupling. For the linear ME coupling this effect is supposed to be similar to the exchange bias effect \cite{MeBe57}, which is sizable for FM nanoparticles.}

\section{Discussion and conclusions}
We consider the superparamagnetic behavior of composite  multiferroics, in particular Fe nano islands on BaTiO$_3$. For realistic material parameters we demonstrated Fe magnetization switching control via external electric fields which should be feasible experimentally by means of telegraph noise. An important finding is that the FM switching times, expressed as mean first passage times \cite{Brow63}, can be tuned by  orders of magnitude (Fig. \ref{fig_3}) by magnetoelectric coupling and electric fields acting on the FE polarization. Our theoretical tools included analytical results and numerical simulations. By this we stress that analytical formulae (e.g. eq. (\ref{eq_11})) should be applied carefully, since the analytical derivations are based on numerous approximations, e.g. on the high energy barrier limit and absorbing boundary conditions, meaning that multiple barrier crossings are not possible. In particular, Fig. \ref{fig_2} shows that as long as the ME coupling is small ($\lambda<0.1$ [s/F]), theory and simulations are in good agreement (cf. dots and curves). It should be mentioned that from telegraph noise experiments performed on such systems one could determine the values of the ME coupling, assuming that it is linear. In this case we expect that fitting of the switching time as a function of temperature according to eq. (\ref{eq_11}) should be realizable. From switching of the polarization direction it should be possible to derive the value of the effective field induced by the ME coupling and finally to relate it to the ME coupling constant from eqs. (\ref{eq_6}) and (\ref{eq_11}). We stress, however, that in general eq. (\ref{eq_11}) is symmetric with respect to the applied $b$-field, therefore, in order to obtain sizable effect experimentally, one should fix the initial magnetization state, e.g. by external applied magnetic field. It should also be emphasized that the so called pinning site effects reported in Refs. \cite{TaSt02,PiLe04,VoBl10} will not strongly modify the  {results} reported here, since the ferroelectricity is controlled by a strong electric field, which indirectly acts on the magnetization.\\
 {Further forms of the ME coupling might strongly modify the obtained results. Assuming that the ME coupling energy has an additional contribution that scales  as $\lambda_1 P_{1 \mathrm{z}}\vec{M}^2$, then the total effective field acting on the magnetization attains  additionally  a term that is linear in magnetization. This would  lead to the modification of the height of the energy barrier defined by the magnetocrystalline anisotropy strength $K_1$. When the magnetization is aligned along the Z-direction, the coupling with $\lambda_1$ will result in a modification of the prefactor $K_1V_{\mathrm{FM}}/(k_{\mathrm{B}}T)$ in eq. (\ref{eq_11}). Further contributions like $\lambda_2 (\vec{P}_1\cdot \vec{M})^2$ would  modify eq. (\ref{eq_11}) in a similar way with a different strength of the net ME coupling.} \\
The final remark concerns the form of the FM nanoparticles, which in general might have a form different from an ideal sphere leading to the effect of sufficiently strong demagnetizing fields \cite{Coey10,OHan00}. The presented results rely on   the idealization  of flat nanoparticles (demagnetizing factor $N_{\mathrm{z}}=1$), giving rise to an axially symmetric contribution to the FM free energy (eq. (\ref{eq_12})) and correcting thus the total effective field acting on the magnetization. Elliptical form of FM nanoparticles (especially with high or low ratio of major ellipsoid axes $c/a$) will lead to a non-axially symmetric FM free energy contribution. Experimentally, following the procedure of Ref. \cite{GoHa13} on Fe/BaTiO$_3$ while lowering Fe coverage our proposal in Fig.\ref{fig_1} should be well within reach.

\section{Acknowledgements}
The authors gratefully acknowledge fruitful discussions with M. Alexe, D. Hesse, and H. Meyerheim on possible experimental realization and the support by grants from the German Research Foundation (Nos. SU 690/1-1 and SFB 762),  {CONACYT of Mexico (Basic Science Projects No. 129269)}, the National Basic Research Program of China (No. 2012CB933101) and the National Natural Science Foundation of China (No. 11104123).


\begin{thebibliography}{40}
\bibitem{footnote_1}
The addition of thermal field to both precesion and relaxation terms of the equation of motion stresses the Landau-Lifshitz-\cite{LaLi35} and not the Kubo-form\cite{Kubo6370} for the spin dynamics.
\bibitem{footnote_3}
We neglect effects of the external electric field on the metallic ferromagnetic part of the system.
\end{thebibliography}
\end{document}